\documentclass[twocolumn,showpacs,preprintnumbers,amsmath,amssymb]{revtex4}

\usepackage{graphicx}
\usepackage{dcolumn}
\usepackage{bm}

\newcommand{\bra}[1]{\langle#1\rvert}
\newcommand{\ket}[1]{\lvert#1\rangle}

\expandafter\ifx\csname
 natexlab\endcsname\relax\def\natexlab#1{#1}\fi
 \expandafter\ifx\csname bibnamefont\endcsname\relax
   \def\bibnamefont#1{#1}\fi
 \expandafter\ifx\csname bibfnamefont\endcsname\relax
   \def\bibfnamefont#1{#1}\fi
 \expandafter\ifx\csname citenamefont\endcsname\relax
   \def\citenamefont#1{#1}\fi
 \expandafter\ifx\csname url\endcsname\relax
   \def\url#1{\texttt{#1}}\fi
 \expandafter\ifx\csname
 urlprefix\endcsname\relax\fi
 \providecommand{\bibinfo}[2]{#2}
 \providecommand{\eprint}[2][]{\url{#2}}

\begin{document}

\title{Quantum storage via refractive index control}

\author{Alexey Kalachev${}^{1,2}$ and Olga Kocharovskaya${}^1$}
\affiliation{${}^1$Institute for Quantum Studies and Department of
Physics, Texas A\&M University, College Station, TX 77843--4242,
USA,\\ ${}^2$Zavoisky Physical-Technical Institute of the Russian
Academy of Sciences, Sibirsky Trakt 10/7, Kazan, 420029, Russia
}%

\date{\today}

\begin{abstract}
Off-resonant Raman interaction of a single-photon wave packet and
a classical control field in an atomic medium with controlled
refractive index is investigated. It is shown that a continuous
change of refractive index during the interaction leads to the
mapping of a single photon state to a superposition of atomic
collective excitations (spin waves) with different wave vectors
and visa versa. The suitability of refractive index control for
developing multichannel quantum memories is discussed and possible
schemes of implementation are considered.
\end{abstract}


\maketitle

\section{Introduction}

During the past decade the optical quantum memories have became
one of the active areas of research in the field of quantum optics
and quantum information (see the Reviews
\cite{LST_2009,HSP_2010,TACCKMS_2010,SAAB_2010}). Such devices are
considered as basic ingredient for scalable linear-optical quantum
computers and efficient quantum repeaters. For practical quantum
information applications, it is necessary to develop memories
which could store quantum states of light with close to 100\%
efficiency and fidelity, and provide long and controllable storage
times or delay-bandwidth products. In this respect significant
experimental progress has been achieved in demonstration of
optical quantum storage using electromagnetically induced
transparency \cite{CMJLKK_2005,DAMFZL_2005,NGPSLW_2007,CDLK_2008},
photon echo induced by controlled reversible inhomogeneous
broadening
\cite{HLALS_2008,HHSOLB_2008,HSHLLB_2009,HLLS_2010,LMRASSG_2010}
or by atomic-frequency comb
\cite{RASSG_2008,AUALWSSRGK_2010,CRBAG_2010,SBWLAHK_2010,UARG_2010,CUBSARG_2010,BGC_2010},
and off-resonance Raman interaction
\cite{ZDJCMKK_2009,RNLSLLJW_2010,RMLNLW_2010}. Optical quantum
memories are usually assumed to store and recall optical pulses,
such as single-photon wave packets, exploiting inhomogeneous
broadened transitions or modulated control fields. In the present
work we suggest one more possibility. By considering quantum
storage based on off-resonant Raman interaction, we show that
manipulation of refractive index in a three-level resonant medium
allows one to store and recall single-photon wave packets without
using inhomogeneous broadening of the atomic transitions or
manipulating the amplitude of the control field. A single-photon
wave packet may be reversibly mapped to a superposition of atomic
collective excitations with different wave vectors, which is
analogues to that of orthogonal subradiant states created in an
extended atomic ensemble \cite{KK_2006}.

As well as providing an interesting possibility for storage, the
refractive index control may also be useful for optimizing
multiplexing regimes of multimode quantum memories, development of
which is important in the prospect of both quantum communication
\cite{CJKK_2007,SRASZG_2007} and computation \cite{TNM_2008}.
Particularly, multimode memories can significantly increase the
quantum communication rate for short storage times. Different ways
of multiplexing has been suggested
\cite{SNRLLSWJ_2008,LRCMKK_2009,ASRG_2009,VSP_2010}, among which
time domain multiplexing is currently most demanded from the view
point of fiber optical communication. Being combined with any
approach to quantum storage mentioned above, the refractive index
control provides an additional degree of freedom for multiplexing
thereby improving capacity of a multimode quantum storage device
or allowing operation in a multichannel regime. Such a
multiplexing method is closely connected with the angular
\cite{SNRLLSWJ_2008} or holographic \cite{VSP_2010} ones since it
also resorts to phase-matching conditions in an extended atomic
ensemble, but it does not exploit different spatial modes of the
field. In effect, the additional multiplexing capacity is based on
the possibility to use frequency and wavelength of the field in a
storage material as independent parameters.

The paper is organized as follows. In Sec. II, we analyze the
storage and retrieval of single-photon wave packets via refractive
index control during off-resonant Raman interaction. In Sec. III,
suitability of refractive index control for developing
multichannel quantum memories is discussed. In Sec. IV, we
consider possible ways of refractive index manipulation and some
implementation issues.

\section{Storage and retrieval of single-photon wave packets}

As a basic model we consider cavity-assisted quantum storage,
which is motivated by the following reasons. First, enclosing an
atomic ensemble in a cavity makes it possible to achieve high
efficiency of quantum storage with optically thin materials. This
may be especially useful for considered off-resonant Raman
interaction since the cross-section of the two-photon transition
is usually small. Second, there is no need for backward retrieval
when optically thin materials are used, which relieves one of
having to perform phase conjugation of the atomic states used for
storage.

We consider a system of $N\gg 1$ identical three-level atoms which
are placed in a single-ended ring cavity and interact with a weak
quantum field (single-photon wave packet) to be stored and with a
strong classical control field (Fig.~1). The atoms have a
$\Lambda$-type level structure, the fields are Raman resonant to
the lowest (spin) transition, and the cavity is resonant to the
quantum field. We assume that the atoms are not moving being
impurities embedded in a solid state material. The interaction
volume is supposed to have a large Fresnel number, which allows us
to take advantage of one-dimensional approximation. The
Hamiltonian of the three-level system in the dipole and rotating
wave approximations is
\begin{figure}
\includegraphics[width=6cm]{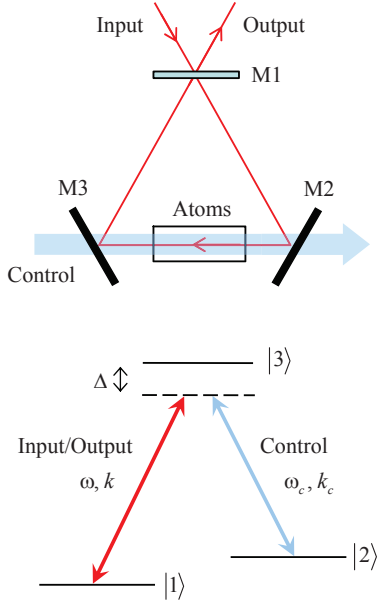}
\caption{\label{fig:levels} (Color online) Schematic of a quantum
memory device (above) and atomic level structure (below). Mirrors
M2 and M3 are perfectly reflecting for the single-photon field and
fully transmitting for the control field, M1 is a partially
transmitting mirror. The difference of wave vectors $k-k_c$ is
modulated via refractive index control during off-resonant Raman
interaction.}
\end{figure}
\begin{equation}
H=H_0+V,
\end{equation}
where
\begin{equation}
H_0=\hbar\omega a^\dag a + \sum_{j=1}^{N}\left(
\hbar\omega_2\sigma_{22}^{j}+\hbar\omega_3\sigma_{33}^{j}\right),
\end{equation}
\begin{equation}
V=-\hbar\sum_{j=1}^{N}\left(\Omega(t)\sigma_{32}^{j}\,e^{ik_cz_j-i\omega_c
t}+ga\sigma_{31}^{j}\,e^{ikz_j}\right)+\text{H.c.}
\end{equation}
Here $\sigma_{mn}^j=\ket{m_j}\bra{n_j}$ are the atomic operators,
$\ket{n_j}$ is the $n$th state ($n=1,2,3$) of $j$th atom with the
energy $\hbar\omega_n$ ($\omega_1=0<\omega_2<\omega_3$), $z_j$ is
the position of the $j$th atom, $a$ is the photon annihilation
operator in the cavity mode, $k_c=\omega_c n_c/c$ and $k=\omega
n/c$ are the wave vectors of the classical and quantum fields,
respectively, $n_c$ and $n$ are refractive indices at the
frequencies $\omega_c$ and $\omega$, $\Omega(t)$ is the Rabi
frequency of the classical field, and $g$ is the coupling constant
between the atoms and the quantized field mode. The values of
$n_c$ and $n$ are considered below as parameters changing in time.
Since variations of them are supposed to be much less than 1, we
leave such time dependence only in phase factors and ignore it in
the factors $\Omega$ and $g$ as functions of refractive indices.

In the Heisenberg picture, we define the following slowly varying
atomic operators: $P_j=\sigma_{13}^{j}{\,e}^{i\omega t}$,
$S_j=\sigma_{12}^{j}{\,e}^{i(\omega-\omega_c)t}$, and cavity field
amplitude $\mathcal{E}=a\,e^{i\omega t}$. From the input-output
relations for the cavity field \cite{CG_1984,GC_1985} we have
\begin{equation}\label{in_out}
\mathcal{E}_\text{out}(t)=\sqrt{2\kappa}\mathcal{E}(t)-\mathcal{E}_\text{in}(t),
\end{equation}
where $2\kappa$ is the cavity decay rate and
$\mathcal{E}_\text{in}$ ($\mathcal{E}_\text{out}$) is the input
(output) field (a single-photon wave-packet). From the
Heisenberg---Langevin equations, assuming that all the population
is in the ground state initially and taking into account that the
quantum field is weak, we find
\begin{align}
\dot{P_j}&=-(\gamma_P+i\Delta)P_j+i\Omega
S_j\,e^{ik_cz_j}+ig\mathcal{E}\,e^{ikz_j},\label{1}\\
\dot{S}_j&=-(\gamma_S+i\Delta_S)S_j+i\Omega^\ast
P_j\,e^{-ik_cz_j},\label{2}\\
\dot{\mathcal{E}}&=-\kappa\mathcal{E}+ig\sum_j
P_j\,e^{-ikz_j}+\sqrt{2\kappa}\,\mathcal{E}_\text{in}.
\end{align}
Here $\gamma_P$ and $\gamma_S$ are the rates of dephasing,
$\Delta=\omega_3-\omega$ is a one-photon detuning, and
$\Delta_S=\omega_2+\omega_c-\omega$ is a two-photon detuning. We
have not included the Langevin noise atomic operators since they
make no contribution to normally ordered expectation values in
consistence with the approximation that almost all atoms remain in
the ground state (see, e.g., \cite{GALS_2007_1} for discussion).
Finally, in the Raman limit, when the single-photon detuning is
sufficiently large, adiabatically eliminating $P_j$ in Eqs.
(\ref{1}) and (\ref{2}), and going to the collective atomic
operators
\begin{equation}
S_q=\frac{1}{\sqrt{N}}\sum_{j} S_{j}{\,e}^{-iqz_j},
\end{equation}
we obtain
\begin{align}
\dot{S}_{q}&=-\gamma'S_{q}+ig'\sqrt{N}\phi(k-k_c-q)\mathcal{E},\label{A1}\\
\dot{\mathcal{E}}&=-\kappa\mathcal{E}+\sqrt{2\kappa}\,\mathcal{E}_\text{in}\nonumber\\
&\qquad\qquad+ig'^\ast\sqrt{N}\sum_{q}\phi(q+k_c-k)S_{q}.\label{A2}
\end{align}
Here $\phi(q)={1}/{N}\sum_{j=1}^N\exp{(iqz_j)}$ is the diffraction
function, $\gamma'=\gamma_S+\gamma_P|\Omega|^2/\Delta^2$,
$g'=g\Omega^\ast/\Delta$, and the resulting frequency shift
$\Delta'=\Delta_S+|\Omega|^2/\Delta$ has been compensated by
tuning the coupling field frequency. The wave vectors $q$ are
multiples of $2\pi/L$, where $L$ is the length of the atomic
medium.

The phase mismatching factors $\phi(q)$, which are usually ignored
on the assumption that a single spatial mode of the spin coherence
is excited and phase-matching is perfect, are now considered.
Suppose that we can manipulate the difference $q+k_c-k$ by
refractive index control without changing frequencies and
propagation directions of the interacting fields. We discuss
possible ways of implementation below. For now it is sufficient to
consider the case when one of the wave vectors, say $k_c$ of the
control field, is changed linearly in time during the interaction
so that $q+k_c-k=(\omega_c/c)\dot{n}_c(t-t_q)$, where $t_q$ is the
moment when $q+k_c-k=0$ for a given $q$. Then
\begin{equation}\label{phi}
\phi[\pm(q+k_c-k)]=e^{\pm
i\beta(t-t_q)}\,\text{sinc}[\beta(t-t_q)],
\end{equation}
where $\text{sinc}(x)=\sin(x)/x$ and
$\beta=({\omega_c}/{c})({L}/{2})\dot{n}_c$. The phase factors
$e^{\pm i\beta t}$ may be compensated by linear or sawtooth phase
modulation of the control field. In such a situation, the phase of
the control field remains constant at the point $z=L/2$ during the
refractive index change. As a result, Eqs.~(\ref{A1}) and
(\ref{A2}) take the form
\begin{align}
\dot{S}_{q}&=-\gamma'S_{q}+ig'\sqrt{N}\,e^{i\beta t_q}\,\text{sinc}(\beta(t-t_q))\,\mathcal{E},\label{A1a}\\
\dot{\mathcal{E}}&=-\kappa\mathcal{E}+\sqrt{2\kappa}\,\mathcal{E}_\text{in}\nonumber\\
&\qquad +ig'^\ast\sqrt{N}\sum_{q}\,e^{-i\beta
t_q}\,\text{sinc}(\beta(t-t_q))\,S_{q}.\label{A2a}
\end{align}

Now it is possible to consider storage and retrieval of a
single-photon wave packet. Let the atomic system interacts with
the quantum field during the time interval $[-T,0]$ with the
initial condition $S_q(-T)=0, \forall q$. Then from Eqs.
(\ref{A1a}) and (\ref{A2a}) we have
\begin{align}
S_q(t)=&ig'\sqrt{N}F_q(-T,t,\mathcal{E})\,e^{i\beta t_q-\gamma't},\label{B1}\\
\dot{\mathcal{E}}(t)=&-\kappa\mathcal{E}(t)+\sqrt{2\kappa}\mathcal{E}_\text{in}(t)\nonumber\\
&-|g'|^2 N\sum_q
\text{sinc}[\beta(t-t_q)]F_q(-T,t,\mathcal{E})\,e^{-\gamma't},\label{B2}
\end{align}
where
$F_q(-T,t,\mathcal{E})=\int_{-T}^t\mathcal{E}(\tau)\,\text{sinc}[\beta(\tau-t_q)]\,e^{\gamma'\tau}
d\tau$. If the cavity field $\mathcal{E}$ varies slowly than
$\delta=\pi/\beta$, and $\gamma'\delta\ll 1$, then Eq.~(\ref{B2})
takes the form
\begin{equation}
\dot{\mathcal{E}}=-\kappa\mathcal{E}+\sqrt{2\kappa}\mathcal{E}_\text{in}-\Gamma\mathcal{E},
\end{equation}
where $\Gamma=|g'|^2N\delta/2$. Then the cavity field can be
adiabatically eliminated provided that $\Gamma+\kappa$ is much
greater than the bandwidth of the input field, which gives
$\mathcal{E}=\sqrt{2\kappa}(\kappa+\Gamma)^{-1}\mathcal{E}_\text{in}$,
and from Eq.~(\ref{B1}) we find
\begin{align}
S_q(0)&=\frac{ig'\sqrt{N}\sqrt{2\kappa}}{\kappa+\Gamma}F_q(-T,0,\mathcal{E}_\text{in})\,e^{i\beta t_q}\nonumber\\
&=\frac{ig'\sqrt{N}\sqrt{2\kappa}}{\kappa+\Gamma}\frac{\pi}{\beta}\mathcal{E}_\text{in}(t_q)\,e^{(i\beta+\gamma')
t_q}.\label{C}
\end{align}
Equation (\ref{C}) describes the mapping of an input single-photon
wave packet to a superposition of collective excitations (spin
waves) with different wave vectors. If the rate of refractive
index change is sufficiently large so that $\delta$ is smaller
than the fastest time scale of the input pulse, then the temporal
shape $\mathcal{E}_\text{in}(t)$ is imprinted on the amplitude
$S_q(0)$ as a function of wave vector $q$. Retrieval is achieved
by off-resonant interaction of the atomic system with the control
field when the values of $n_c$ that used for storage are scanned
again. Suppose, e.g., that the time dependence of refractive index
during the time interval $[0,T]$, when
$\mathcal{E}_\text{in}(t)=0$, is reversed. In this case, instead
of Eq.~(\ref{B2}) we have
\begin{align}
\dot{\mathcal{E}}(t)&=-\kappa\mathcal{E}(t)\nonumber\\
&\quad-\frac{|g'|^2N\sqrt{2\kappa}}{\kappa+\Gamma}\sum_q
\text{sinc}[\beta(t+t_q)]F_q(-T,0,\mathcal{E}_\text{in})\,e^{-\gamma't}.
\end{align}
For slow-varying $\mathcal{E}_\text{in}$ this equation takes the
form
\begin{equation}
\dot{\mathcal{E}}(t)=-\kappa\mathcal{E}(t)-\frac{\Gamma\sqrt{2\kappa}}{\kappa+\Gamma}\mathcal{E}_\text{in}(-t)\,e^{-2\gamma't}.
\end{equation}
Finally, after adiabatic elimination of the cavity field and using
Eq.~(\ref{in_out}) we obtain
\begin{equation}\label{D}
\mathcal{E}_\text{out}(t)=-\frac{2\Gamma}{\kappa+\Gamma}\mathcal{E}_\text{in}(-t)\,e^{-2\gamma't}.
\end{equation}
The solution Eq.~(\ref{D}) is exactly the same as that in the
cavity-assisted storage with inhomogeneous broadening
\cite{AS_2010,MAG_2010}. The output field becomes time-reversed
replica of the input field provided that the duration of wave
packet is much smaller that the decay time $1/\gamma'$, and  the
efficiency of the storage followed by retrieval is maximum under
impedance-matching condition $\kappa=\Gamma$. The only difference
is the collective absorption/emission rate $\Gamma$, which in our
case takes the form
\begin{equation}\label{Gamma}
\Gamma=\frac{g^2N|\Omega|^2}{\Delta^2}\frac{\pi}{2\beta}.
\end{equation}
It means that the time interval $\delta=\pi/\beta$, which is
actually the time interval between two adjacent orthogonal spin
states created upon refractive index control, is analogous to
inhomogeneous life-time, i.e., reversal inhomogeneous linewidth.
We see that a single-photon wave packet can be effectively stored
and reproduced via refractive index control in a three-level
system without inhomogeneous broadening and without modulating the
Rabi frequency of the control field during the interaction. It is
also important that the time dependence of the refractive index
need not be reversed during the retrieval. If the values of $n_c$
during retrieval are ordered like those during storage,
Eq.~(\ref{D}) is replaced by
\begin{equation}\label{D1}
\mathcal{E}_\text{out}(t)=-\frac{2\Gamma}{\kappa+\Gamma}\mathcal{E}_\text{in}(t-T)\,e^{-\gamma'T}.
\end{equation}
Thus a single-photon wave packet may be reconstructed without time
reversal so that its temporal shape be not deformed by the
rephasing process.

The total change of refractive index during storage or retrieval
is
\begin{equation}
\Delta
n=\dot{n}_cT=\left(\frac{T}{\delta}\right)\frac{\lambda}{L},
\end{equation}
where $\lambda=2\pi c/\omega_c$. Numerics show that a Gaussian
pulse with the duration (FWHM) as short as $2\delta$ can be stored
and recalled with the efficiency 0.99 provided that $\gamma'$ is
small enough (Fig.~2). Therefore, taking $T/\delta\sim 1$ and
$\lambda/L\sim 10^{-5}$ we have $\Delta n\sim 10^{-5}$, which may
be considered as the minimum refractive index increment needed for
storage of a single pulse under typical experimental conditions.
The ratio between the total accessible range of refractive index
change and this minimum value determines the number of pulses can
be stored in a series, i.e., the mode capacity of quantum memory.
\begin{figure}
\includegraphics[width=6cm]{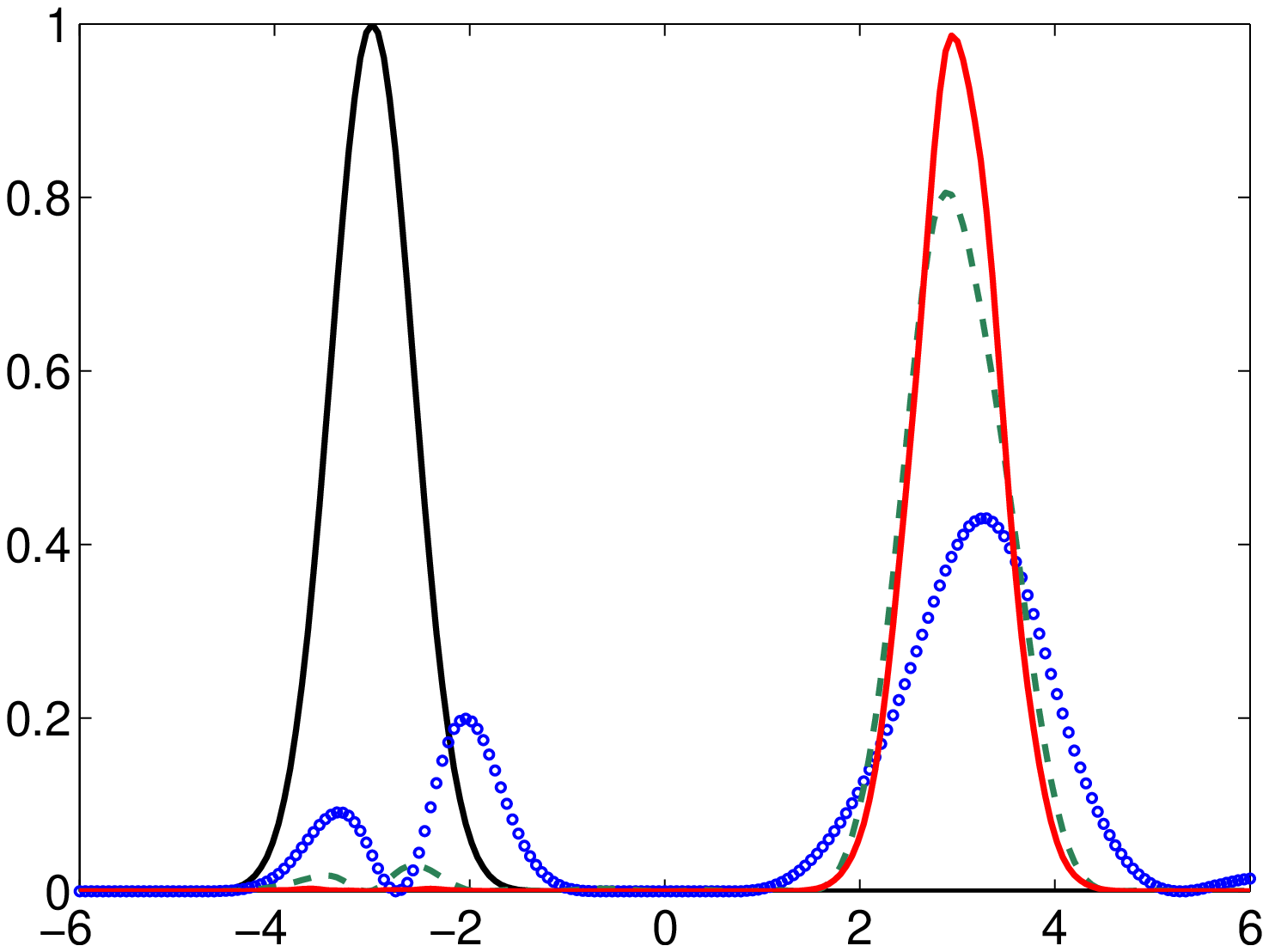}
\put(10,0){$t/\tau_p$} \put(-200,120){$|\mathcal{E}(t)|^2$,}
\put(-200,110){a.u.} \put(-150,110){Input}\put(-150,100){Pulse}
\caption{\label{fig:levels} (Color online) Storage and retrieval
of a Gaussian pulse for different values of $\delta=\pi/\beta$.
The black solid curve is the input pulse of duration $\tau_p$.
Other curves represent the output field for the cases
$\delta=2\tau_p$ (blue circles), $\delta=\tau_p$ (green dashed
line), $\delta=\tau_p/2$ (red solid line). These curves were
obtained by numerically solving Eqs.~(\ref{A1a}) and (\ref{A2a}),
treated as complex number equations, with the condition
$\kappa=\Gamma$ and $\gamma'\tau_p\ll 1$. The sign of $dk_c/dt$ is
changed at the moment $t=0$.}
\end{figure}

The impedance matching condition $\Gamma=\kappa$ may be written in
the following form:
\begin{equation}\label{IM}
C\gamma'\delta/2=1,
\end{equation}
where $C=g^2N/\kappa\gamma=|g'|^2N/\kappa\gamma'$ is the
cooperativity parameter, which could also be expressed in terms of
the cavity finesse $\mathcal{F}$ and resonant absorption
coefficient $\alpha$ as $C=\alpha L\mathcal{F}/2\pi$, provided
that the resonant medium fills the cavity. Thus to store and
recall pulses broader than the Raman linewidth $\gamma'$, we need
to increase $C$ above 1 appropriately. In general case, $\gamma'$
consists of homogeneous and inhomogeneous contributions. If we do
nothing with the latter, the delay-bandwidth product of quantum
storage (or multimode capacity) proves to be of the order of $C$.
It may be increased by choosing materials with a narrower
line-width or shorter pulses with appropriate increasing the
cooperative parameter. But if we are able to reverse the
inhomogeneous broadening, the delay-bandwidth product may be
increased significantly depending on a residual noncontrollable
broadening, say $\gamma'_\text{hom}$, of the Raman transition. As
a result, the mode capacity of quantum storage may be $C$ times
larger than that achievable without refractive index control,
which is determined by the ratio $\gamma'/\gamma'_\text{hom}$.

\section{Multichannel quantum storage}

In the previous Section, we were interested in the situation when
the bandwidth of the photon is larger than the inhomogeneous
linewidth of the Raman transition. Now we consider the opposite
case, when the storage and retrieval are implemented, for example,
by manipulating the inhomogeneous broadening, which should be
larger than the photon bandwidth, and refractive index control is
used for realizing multichannel regime of quantum storage. The
idea is that different wave vectors of the spin waves correspond
to different channels of storage and retrieval just as in the case
of angular multiplexing. For example, consider the Raman echo
memory scheme with controlled reversible inhomogeneous broadening
that is switched in time periodically \cite{HSHLLB_2009}. The
multichannel regime can be achieved by assigning different wave
vectors $k_c$ to the control field (in our case --- by refractive
index control) during different dephasing/rephasing circles. In a
similar way, we can consider memory schemes based on resonant
interaction. Let the storage and retrieval be implemented using
atomic frequency comb \cite{ASRG_2009}, which dephases and then
rephases after a time $T$, and $\pi$-pulses transferring the
optical coherence to/from the spin coherence are used
\cite{AUALWSSRGK_2010}. Then we can make different $n$ for
different $\pi$-pulses thereby creating orthogonal spin waves on
different storage/retrieval cycles or we can change refractive
index for the weak field to be stored so that $\delta=T$, which
leads to the same result. In any case, such multichannel regimes
enable one to process new quantum states while preserving those
stored before and provides access to all states kept in store in
any order. It is also important that the phase modulation of the
control field, which is required to store and recall pulses
without resort to inhomogeneous broadening, becomes needless in
the case of such channel division. Regarding the impedance
matching condition, it takes the form $C=1$ since $\delta/2$ is
replaced by inhomogeneous life-time $T_2^\ast=1/\gamma'$.

Apart from the finite storage time due to irreversible relaxation,
the maximum number of channels is also limited by the
signal-to-noise ratio. The latter can be estimated in the
following simple way. Consider two channels corresponding to wave
vectors with the difference $\Delta k_m=2\pi m/L$, where $m$ is an
integer. During retrieval from one of them the probability of
retrieval from another is determined by phase mismatching and
proportional to $P_m=[\text{sinc}(\Delta k_mL/2)]^2$, which may be
equal to zero only for monochromatic field. Now it is necessary to
take into account the bandwidth $\delta k=\delta\omega n/c$ of the
retrieved signals (here $n$ is the average value of refractive
index during storage or retrieval). If $\delta k\ll k$, then
\begin{equation}
P_m\approx\frac{1}{\delta k}\int_{-\delta k/2}^{\delta
k/2}\left[\frac{Lx}{2\pi
m}\right]^2\,dx=\frac{1}{12}\left[\frac{\delta\omega}{\omega}\frac{L}{\lambda}\frac{n}{m}\right]^2.
\end{equation}
We see that the noise from another channel is quadratically
proportional to the bandwidth of the photons $\delta\omega$ and
length of the sample $L$. As an example, let $L/\lambda= 10^5$,
$\omega/2\pi=2\cdot 10^{14}\;\text{Hz}$, and $n=2$. If
$\delta\omega/2\pi\sim 50\;\text{MHz}$, then $P_m\lesssim 10^{-4}$
for $m\geq 1$ so that 100 channels provide total signal-to-noise
ratio of the order of 100. In order to maintain this ratio for
broader signals, the value of $m$ should be increased
proportionally to the bandwidth, which means increasing refractive
index difference between adjacent channels.

\section{Refractive index control}

Let us discuss possible ways of refractive index manipulation that
are suited to quantum storage devices. First, if a doped nonlinear
crystal is used as a storage medium, we can take advantage of the
linear electro-optic effect. For example, the maximum value of the
index change in $\text{LiNbO}_3$, which is limited by the
breakdown electric field, is of the order of $10^{-3}$. Although
attaining this value by applying a moderate voltage is possible
only for a waveguide configuration, the doped nonlinear materials,
particularly $\text{LiNbO}_3\text{:Er}^{3+}$, hold promise in
quantum storage applications, and therefore traditional
electro-optic techniques of refractive index manipulation might be
of helpful. Second, the resonant enhancement of the refractive
index with vanishing absorption can be realized via quantum
interference effects
\cite{S_1991,FKSSUZ_1992,RFZHS_1993,H_1994,LYZS_1999,Y_2005}.
Although the maximum enhancement demonstrated experimentally in a
gaseous medium is $10^{-4}$ \cite{ZLHNSRV_1996,PUGY_2008}, there
is good reason to expect much larger values, such as $10^{-2}$,
for solid state materials \cite{CBS_2003,YJY_2006}. It should be
noted that manipulating refractive index either on the frequency
of the quantum field to be stored or on the frequency of the
strong control field should not only guarantee small losses but
also no amplification. One way of avoiding the gain is to use
excited state absorption \cite{BK_2009}. Finally, we would like to
note that some possibilities of refractive index control are
provided even by a frequency shift of an absorption structure
relative to the $\Lambda$-type one. At cryogenic temperatures
optical transitions of impurity crystals and especially of
rare-earth ion-doped ones have very narrow homogeneous lines so
that in the neighborhood of an inhomogeneous profile there might
be rather strong frequency dependence of refractive index and yet
very small absorption. In this respect, spectral hole-burning
techniques can be very useful for preparing inhomogeneous profiles
with sharp edges as well as for creating $\Lambda$-systems on a
nonabsorbing background. A more detailed analysis of such an
approach, which invokes specific information about resonant
materials, will be presented elsewhere.

\section{Conclusion}
It is shown that single-photon wave packets can be stored and
recalled in a resonant three-level medium by means of refractive
index control without recourse to inhomogeneous broadening and
modulating the amplitude of control fields. Such a scheme for
quantum storage can be combined with other techniques to increase
the mode capacity of quantum memories or to develop multichannel
storage devices. Although implementation of refractive index
control is still a challenging experimental problem, we hope that
significant progress can be made in this field by taking advantage
of solid-state materials, which are also promising for quantum
storage applications.

%


\end{document}